# Does a standalone, "cold" (low-thermal-noise), linear resistor exist without cooling?

Jiaao Song and Laszlo B. Kish [†]

*Department of Electrical and Computer Engineering, Texas A&M University, TAMUS 3128, College Station, TX 77843-3128*

**Abstract.** Classical ways of cooling require some of these elements: phase transition, compressor, non-linearity, valve, and/or switch. A recent example is the 2018 patent of Linear Technology Corporation; they utilize the shot noise of a diode to produce a standalone nonlinear resistor that has *T*/2 noise temperature (about 150 Kelvin). While such "resistor" can cool its environment when it is AC coupled to a resistor, the thermally cooling effect is only academically interesting. The importance of the invention is of another nature: In low-noise electronics, it is essential to have resistors with low noise temperature to improve the signal-to-noise ratio. A natural question is raised: can we use a linear system with feedback to cool and, most importantly, to show reduced noise temperature? Exploring this problem, were able to produce standalone linear resistors showing strongly reduced thermal noise. Our must successful test shows *T*/100 (about 3 Kelvin) noise temperature, as if the resistor would have been immersed in liquid helium. We also found that there is an old solution offering similar results utilizing the virtual ground of an inverting amplifier at negative feedback. There the "cold resistor is generated at the input of an amplifier. On the other hand, our system generates the "cold" resistance at the output, which can have practical advantages.

## 1. Introduction

Resistors produce thermal noise that can be reduced by cooling. However, cooling is energy needy, expensive and bulky therefore there is a high technological interest to imitate a resistor with lower absolute temperature than its environment.

*1.1. Thermal noise in resistors and the Second Law*

The thermal noise (Johnson-Nyquist noise) of resistors is due to the thermal motion of charge carriers in the sample, and it is the manifestation of Boltzmann's energy equipartition theorem. Its general existence in the quantum (high-frequency/low-temperature) limit is debated [1,2]. However the present paper is about its classical (low-frequency/high-temperature) limit, where there is a common agreement that the power density spectrum $S_u(f)$ of the thermal noise voltage of a resistor of resistance *R* is correctly given by the Johnson-Nyquist formula used in electrical engineering [3]:

$$S_u(f) = 4kTR \quad , \tag{1}$$

where *k* is the Boltzmann constant and *T* is the temperature of the resistor in thermal equilibrium. In a special non-equilibrium situation, Equation 1 and the linear response theory implies that the electrical power flow between two parallel-connected resistors with different temperatures (see Figure 1), in the frequency band $\Delta f$, is given as [4]:

---

[†] Corresponding Author

$$P_{h \to c} = 4k(T_h - T_c)\frac{R_h R_c}{(R_h + R_c)^2}\Delta f \ , \qquad (2)$$

where $P_{h \to c}$ is the mean power flow from the "hotter" resistor $R_h$ of temperature $T_h$ to the "colder" resistor $R_c$ of temperature $T_c$.

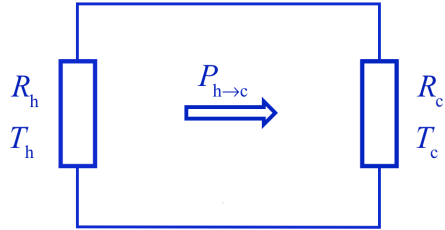

**Figure 1**. Thermo-electric power flow between two parallel resistors [4].

When the circuitry in Figure 1 is in thermal equilibrium, the net power flow between two resistors is zero as implied by the Second Law of Thermodynamics. Conversely, at a fixed ambient temperature, in thermal equilibrium, the thermal noise of a resistor cannot be reduced without incorporating an active device [5] that will dissipate enough to avoid Second Law violation. When such resistor is kept at the same temperature as its environment, it would act as a cooler: it would extract heat from the circuitry it is connected to.

*1.2 Reducing thermal noise in electronics*

Of course, the physical cooling aspects of reduced thermal noise is typically marginal and the issue is important for another reason: the research and development of low-noise electronics. It is widely taught [3] that, in accordance with Equation 1, the only way to reduce the thermal noise of a given resistor is the lowering its temperature. Such applications indeed exist, for example in radio astronomy and space applications, however for a great price. Cooling is bulky, expensive and extremely energy-needy. This situation motivates the perplexing question:

*Is it possible to develop an active linear circuitry that, without a cooler, would be able to imitate the response and thermal noise of a standalone linear resistor that is colder than the ambient temperature?*

*If the answer is yes, such a circuitry could be used as a "plug-in" substitute for resistors at sensitive places where their thermal noise is an issue in order to produce record-low-noise electronics without expensive, bulky and energy needy cryogenic cooling.*

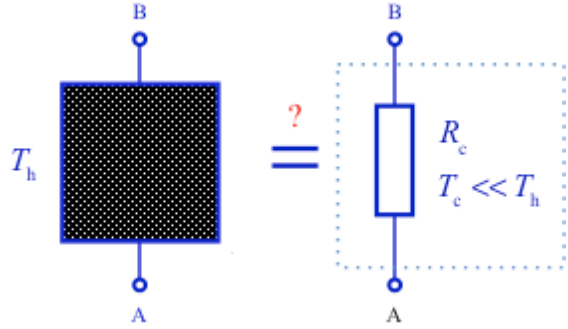

**Figure 2**. "Cold" resistor (active circuitry) in a black box.

Moreover, such system would be also academically interesting, at least for statistical physicists, particularly if the system would be truly linear and the temperature reduction were more than 50%. Before we show our own concept, to illustrate the practical interests about the matter, we cite two types of earlier achievements: the diode as resistor for small signals, and resistors embedded in a feedback circuitry.

*1.3 The diode as a "cold" resistor*

The high technological interest to imitate a resistor with lower absolute temperature than the ambient temperature is indicated by a recent patent [6] of Linear Technology Corporation, where they utilize the shot noise of a diode to produce an AC-coupled nonlinear resistor that has $T/2$ noise temperature when its differential resistance is used in Equation 1, see Figure...

The diode is forward-driven by a low-noise DC current generator and the AC coupling is done by the capacitor. For small voltages $<< kT/q$, where q is the elementary charge, this "system in a black box" imitates a linear resistor with resistance equal to the differential resistance $kT/qI$ of the diode and its thermal noise with $T/2$ temperature.

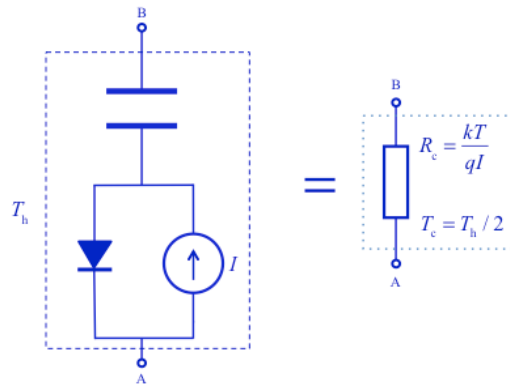

**Figure 3**. Forward biased, AC-coupled diode in a black box as cold ($T/2$) resistor in the small-signal limit [6].

The disadvantage of this system that it is basically a non-linear system [7] thus it can tolerate very small, $<< kT/q$, signals only; moreover the 1/2 temperature reduction factor is not really impressive; though it is much more than nothing.

*1.4 Active resistors with seemingly lowered temperature*

A standalone, linear, active resistor with decreased noise temperature could be used as a "cold resistor plugin" in general electronic applications and could make expensive cryogenic cooling unnecessary in some cases and for some parts of ultra-low-noise electronics. Naturally, it could outperform and make the diode-based patent obsolete in many situations. In the next section, we describe our solution of this challenge. In Sections 2 and 3, we show our solution and in Section 4, we will discuss a different method obtained by earlier others.

**2. "Standalone" resistor "plugin" with lowered temperature**

Going beyond the efforts of [8,9], we present the following challenge:

(i) Can we construct a *linear* circuitry in a black box at physical temperature $T_h$, where the black box behaves as a standalone linear resistor with reduced temperature of $T_c = rT$ with $r<1$ ?

(ii) In other words: Can we cool by a linear amplifier without non-linearity, switches, or phase transition?

(iii) If the answer is "yes", can we go below $r = 0.5$, which is the lower limit for an ideal diode, a *nonlinear* element [3] ? Can we reach $r << 0.5$, and beat the Linear Technology Corporation patent [3] ?

Below we introduce the solution that can provide "yes" answer to questions (i),(ii) and that positively answers question (iii) by offering $r << 0.5$ (see the demonstration with r=0.01, that is, 3 Kelvin effective temperature in Section 3). To help the reader and the logic flow, first we present some simple preparations; some of them are well-known.

*2.1 Impact of negative feedback on a voltage generator at the output of linear amplifiers*

For these considerations, see the left part of Figure 4 where a voltage generator is placed at the output (point D) of an ideal, linear amplifier with amplification $A_1$. The new output is then point B. From point B, another linear amplifier with amplification $A_2$ executes a feedback to the input (point C) of the first amplifier. We assume that the amplification of one of the amplifiers is negative, while that of the other one is positive, thus the loop-amplification $A_1 A_2$ is negative:

$$A_1 A_2 < 0 \tag{3}$$

In the ideal case (linear, noise-free, zero output impedance, infinite bandwidth), the voltages at the different points in Figure 4 are interrelated as follows:

$$U_D = A_1 U_C \tag{4}$$

$$U_C = A_2 U_B \tag{5}$$

$$U_\text{B} = U_\text{D} + U_0 \tag{6}$$

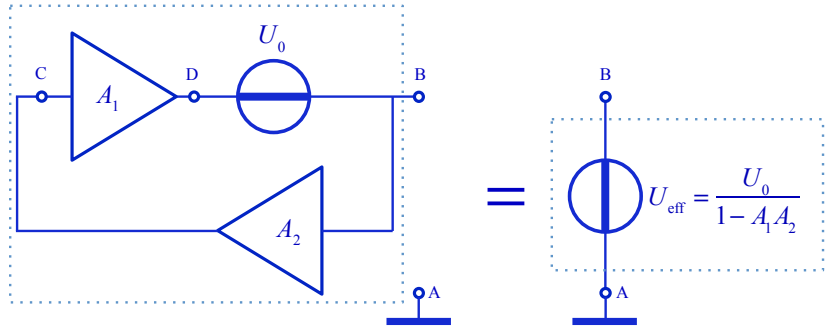

**Figure 4.** Reducing an additive voltage at the amplifier output by negative feedback (left) and the equivalent circuit (right). Note, in the case of negative feedback $A_1 A_2$ has negative value.

Solving Equations 4-6, we obtain that the output voltage is the system is a reduced value of the $U_0$ generator voltage:

$$U_\text{B} = \frac{U_0}{1 - A_1 A_2} . \tag{7}$$

From Equations 3,7, we obtain:

$$U_\text{B} = \frac{U_0}{1 + |A_1 A_2|} . \tag{8}$$

Therefore, the circuitry on the left in Figure 4 is equivalent to the grounded voltage generator on the right with reduced voltage:

$$U_\text{eff} = \frac{U_0}{1 + |A_1 A_2|} . \tag{9}$$

Note, even though Equation 8 is typically not in textbooks, its inherent notion is widely used, when negative feedback is utilized to decrease the distortion, increase the bandwidth and decrease the output impedance of practical amplifiers.

*2.2 Impact of negative feedback on the output resistance of linear amplifiers*

Figure 5 (left) shows a different situation: a serial resistor $R_0$ at the output of the amplifier with the same negative feedback as in Figure 4. The equivalent circuitry is well-known and it is a grounded resistor with reduced resistance:

$$R_\text{eff} = \frac{R_0}{1 + |A_1 A_2|} . \tag{10}$$

A simple explanation is to assume that a current generator feeds the output at point B. Due to Ohm's law the current will generate a voltage on the resistor. Due to the feedback, the application of Equation 9 will lead to a reduced voltage and a reduced effective resistance by the same factor in Equations 9 and 10, respectively.

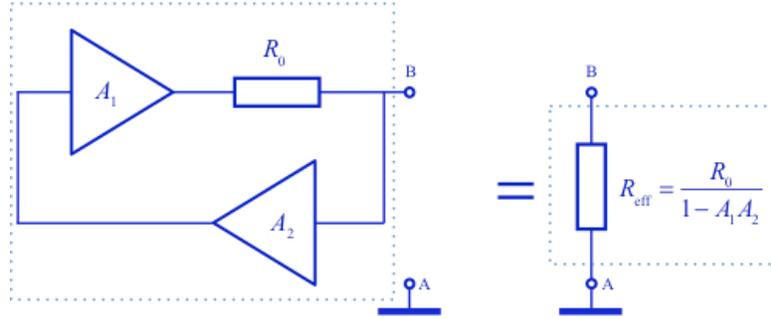

**Figure 5**. Reduction of the output resistance by negative feedback ($A_1A_2$ has negative value).

*2.3 Impact of negative feedback on a resistance at the output and its thermal noise*

Combining the results in subsections 2.1 and 2.2, the results are straightforward and shown in Figure 6. Both the resistance value and the thermal noise voltage are reduced by the same factor as given above. Because the noise spectrum $S(f)$ in Equation 1 is scaling with the square of the effective noise voltage, the noise spectrum is reduced by

$$S_{\text{eff}}(f) = \frac{S(f)}{\left(1+|A_1A_2|\right)^2} \quad . \tag{11}$$

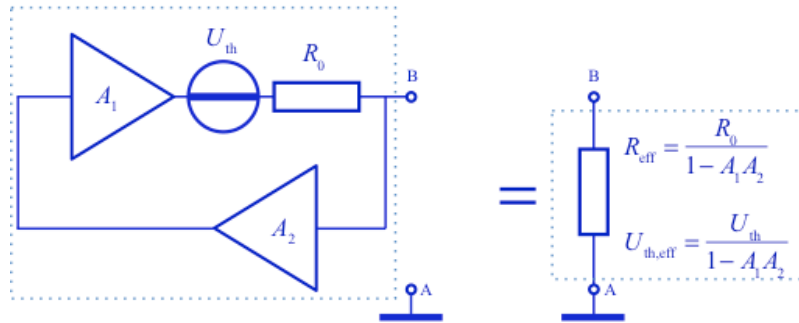

**Figure 6**. Reduction of the resistance and thermal noise of a resistor at the output, in the case of negative feedback ($A_1A_2$ has negative value).

*2.4 Standalone "cold" resistor with reduced temperature*

From Equations 1 and 11, it is obvious that the effective resistor represented by the circuit in Figure 6, in the case of negative feedback ($A_1A_2$ has negative value) has a resistance and temperature reduced by the same factor:

$$T_c = \frac{T_h}{1+|A_1A_2|} \quad . \tag{12}$$

In an ideal case, the achievable temperature reduction would have no lower limit, however, in practical cases, there are limitations imposed by non-ideal features, see the achieved limit below.

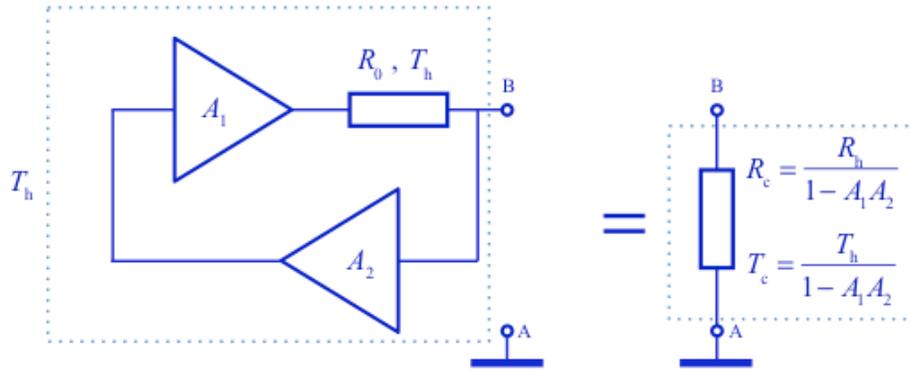

**Figure 7**. Active resistor "plugin" with strongly reduced noise temperature. Note, $A_1A_2$ has negative value (negative feedback.

**3. Test results with practical integrated circuits**

Real circuits have non-zero input current and voltage noises; finite bandwidth; phase dispersion; non-linearity, and non-zero output resistance. Therefore the practical behavior will follow the results of the ideal model presented in section 2 in those situations only when the deviations caused by the above-listed non-ideal features impose negligible effects on the performance.

We used the *LTspice XVII* industrial circuit tester/simulator system to measure the practical performance of the temperature reduction scheme proposed in Section 2.

In Figure 8 the circuit representation of the "cold" resistor plugin of Figure 7 is shown. The equivalent cold linear resistor with resistance $R_{\text{eff}}$ and noise temperature $T_c$ is shown in the right. $LA_1$, $LA_2$, $LA_3$, are ADA4627 operational amplifiers. $R_0$ = 500MOhm, $R_1=R_3$=1kOhm. $R_2$ was varied to control $R_{\text{eff}}$ and $T_c$.

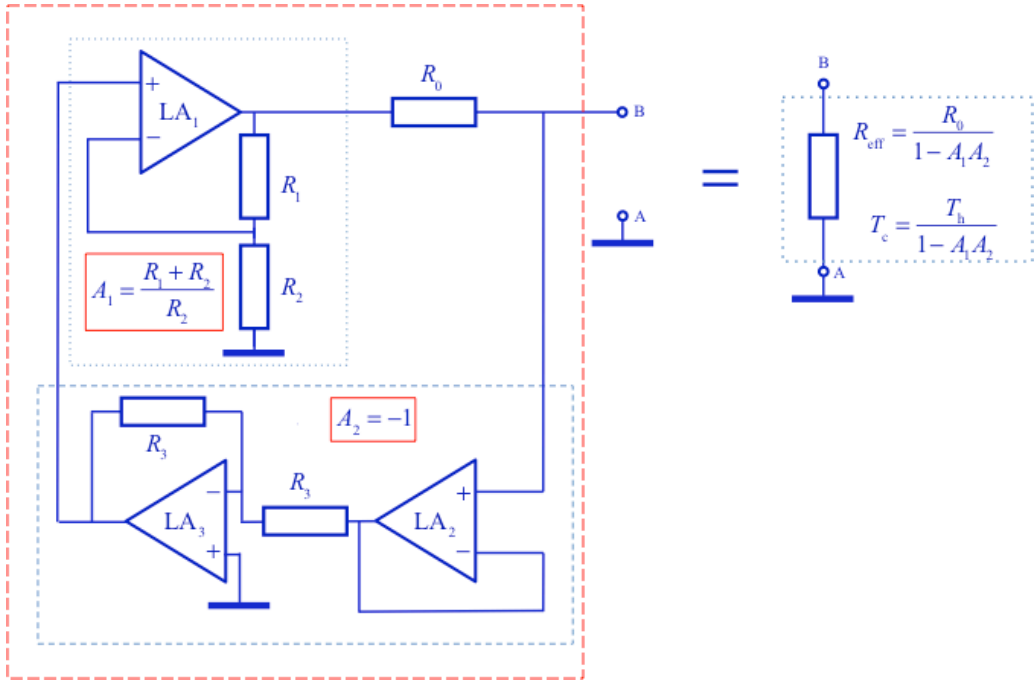

**Figure 8**. Left: Circuit representation of the "cold" resistor plugin of Figure 7. Right: equivalent cold linear resistor with resistance $R_{eff}$ and noise temperature $T_c$. $LA_1$, $LA_2$, $LA_3$, are ADA4627 operational amplifiers. $R_0$ = 500MOhm, $R_1$=$R_3$=1kOhm. R2 was varied to control $R_{eff}$ and $T_c$.

Figure 9 shows the projected ideal temperature reduction factor versus $R_2$. The practical, effective reduction is limited by non-ideal features such as the noise of the amplifiers and resistors, phase shift, frequency characteristics, etc.

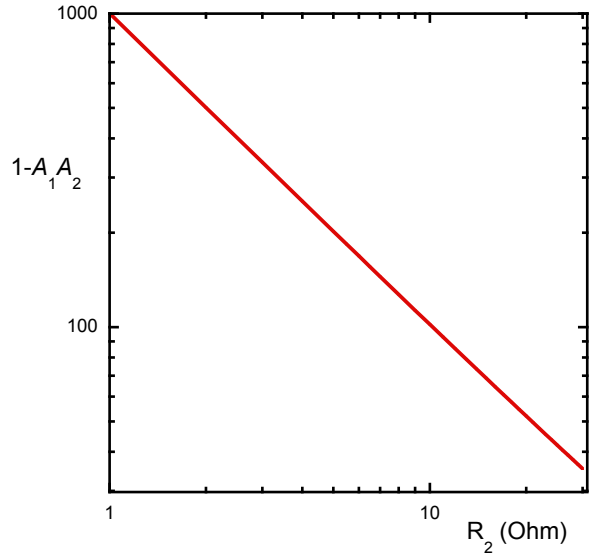

**Figure 9**. The ideal temperature reduction factor versus $R_2$. The practical, effective reduction is limited by non-ideal features such as the noise of the amplifiers and resistors, phase shift, frequency characteristics, etc.

Figure 10 shows the practical effective temperature $T_c$ and resistance $R_{eff}$, measured at 1 kHz frequency, with the circuitry in Figure 8. $R_{eff}$ was measured by forcing a probe current to point B and evaluating the voltage response. $T_c$ was measured by measuring the noise voltage spectrum $S_u$ at point B and evaluating the noise temperature by Equation 1 using the measured value of $R_{eff}$.

The lowest effective temperature $T_c$ was about 3 Kelvin, which is liquid Helium temperature and it is 50 times lower than the limit the Linear Technology Corporation patent can reach with a diode.

Note, we have not optimized our system for the noises of the resistors, and phase and frequency characteristics. The different ways of deviation of $T_c$ and $R_{eff}$ from the ideal value indicate that with proper optimization a significantly lower temperature can be reached. This question is an object of future research.

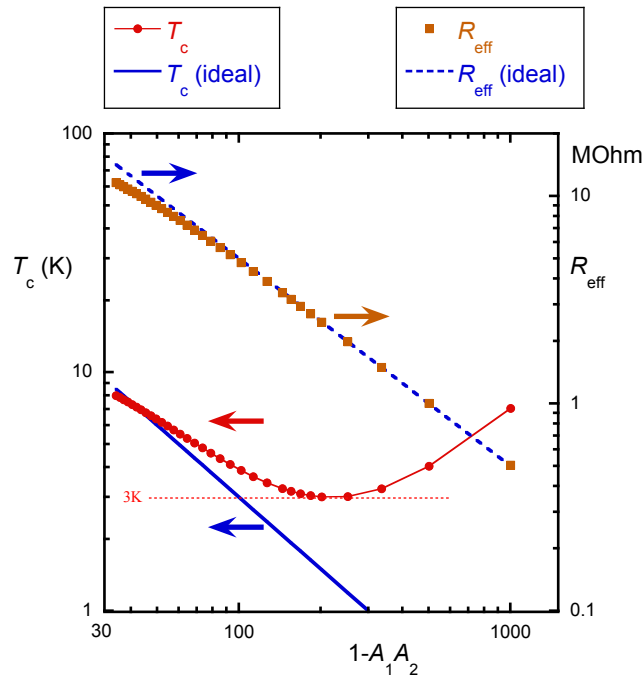

**Figure 10**. The practical effective temperature $T_c$ and resistance $R_{eff}$ measured at 1 kHz frequency at the circuit shown in Figure 8. $R_{eff}$ was measured by forcing a probe current to point B and evaluating the voltage response. $T_c$ was measured by measuring the noise voltage spectrum $S_u$ at point B and evaluating the noise temperature by Equation 1 using the measured value of $R_{eff}$. The lowest effective temperature $T_c$ was about 3 Kelvin, which is liquid Helium temperature and it is 50 times lower than the limit the Linear Technology Corporation patent can reach with a diode. We have not optimized our system for the noises of the resistors and phase and frequency characteristics of the system. The different types of deviation of $T_c$ and $R_{eff}$ from the ideal value indicate that with proper optimization a significantly lower temperature can be reached. This question is an object of future research.

**4. A common different solution in earlier works**

We note here that after we formed our solution, it came to our attention that, in 1942, Strutt and van der Ziel [8] proposed a different solution to produce "cold" resistors that was followed up in modern applications later [9-11]. They also developed an active resistor by negative feedback however that active resistor was the *input* resistance (impedance) of an amplifier, see Figure 11.

A simple calculation [9] results in similar results as we have obtained about our new system except that amplification $A_1$ is substituted for $A_1 A_2$.

Its disadvantage is that the voltage on the active resistor (between points A and B) appears between the input electrodes of an amplifier with a large amplification ($A_1$). In our new solution described in Sections 2 and 3, see Figures 7 and 8, where the active resistor is the output resistance of an amplifier, this situation can be easily avoided. In the generic situation (Figure 7), the amplification of $A_2$ can be chosen a low value, for example its value is 1 in our demonstration (Section 3). Moreover the LA$_2$ amplifier, which has its input at active point B (see Figure 8), operates in the follower mode that can handle large voltages.

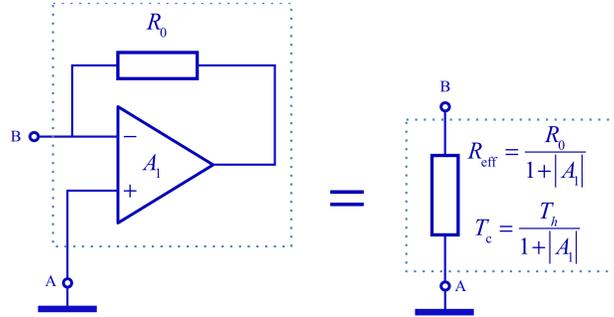

**Figure 11.** The common principle of earlier solutions. Its disadvantage is that the voltage on the active resistor (between points A and B) appears between the input electrodes of an amplifier with a large amplification ($A_1$). In the new solution described in Sections 2 and 3, see Figures 7 and 8, this situation can be avoided.

**Conclusions**

Similarly to [9] we have shown that, from an "academic point of view", it is possible to cool with an active linear system. That means cooling without the usage of phase transitions, pumps, vales, switches or non-linearity. At the practical side, we introduced and demonstrated a linear system that can imitate a low-temperature linear resistor with cryogenic temperature while the circuitry is at room temperature. It can be used as a plugin "cold" resistor in low-noise applications. The method utilizing the active output resistance of an amplifier can, at certain situations, be more advantageous than the earlier ones utilizing active input resistance.

**Acknowledgements**